\documentclass[pra,aps,amsmath,amssymb,showpacs,showkeys]{revtex4}

\newcommand{\hh}{{\mathcal{H}}}
\newcommand{\lnp}{{\mathcal{L}}}
\newcommand{\lsa}{{\mathcal{L}}_{s.a.}}
\newcommand{\lsp}{{\mathcal{L}}_{+}}
\newcommand{\bro}{{\boldsymbol{\rho}}}
\newcommand{\clb}{{\mathcal{B}}}
\newcommand{\cla}{{\mathcal{A}}}
\newcommand{\cmb}{{\mathbb{B}}}
\newcommand{\mma}{{\mathbb{A}}}
\newcommand{\mmb}{{\mathbb{M}}}
\newcommand{\am}{{\mathsf{A}}}
\newcommand{\bn}{{\mathsf{B}}}
\newcommand{\ax}{{\mathsf{X}}}
\newcommand{\ay}{{\mathsf{Y}}}
\newcommand{\az}{{\mathsf{Z}}}
\newcommand{\nq}{{\mathsf{Q}}}
\newcommand{\mm}{{\mathsf{M}}}
\newcommand{\nm}{{\mathsf{N}}}
\newcommand{\vm}{{\mathsf{V}}}
\newcommand{\mc}{{\mathcal{M}}}
\newcommand{\nc}{{\mathcal{N}}}
\newcommand{\pen}{\openone}

\newcommand{\tr}{{\mathrm{tr}}}

\newcommand{\iu}{{\mathtt{i}}}
\begin{document}
\clearpage
\preprint{}

\title{Fine-grained uncertainty relations for several quantum measurements}
\author{Alexey E. Rastegin}
\email{alexrastegin@mail.ru}
\affiliation{Department of Theoretical Physics, Irkutsk State University,
Gagarin Bv. 20, Irkutsk 664003, Russia}

\begin{abstract}
We study fine-grained uncertainty relations for several quantum
measurements in a finite-dimensional Hilbert space. The proposed
approach is based on exact calculation or estimation of the
spectral norms of corresponding positive matrices. Fine-grained
uncertainty relations of the state-independent form are derived
for an arbitrary set of mutually unbiased bases. Such relations
are extended with a recent notion of mutually unbiased
measurements. The case of so-called mutually biased bases is
considered in a similar manner. We also discuss a formulation of
fine-grained uncertainty relations in the case of generalized
measurements. The general approach is then applied to two
measurements related to state discrimination. The case of three
rank-one projective measurements is further examined in details.
In particular, we consider fine-grained uncertainty relations for
mutually unbiased bases in three-dimensional Hilbert space.
\end{abstract}


\maketitle

\pagenumbering{arabic}
\setcounter{page}{1}

\section{Introduction}\label{sec1}

The uncertainty principle \cite{wh27} is one of fundamental
limitations in description of the quantum world. Studies of such
limitations are important in own rights as well as for
applications in quantum information theory. Many quantitative
formulations of the uncertainty principle were proposed. The
Robertson formulation \cite{robert} gives a lower bound on product
of the standard deviations of two quantum observables. The authors
of \cite{mc12} gave a reformulation dependent on the degree of
Gaussianity of the state. The entropy-based approach is an
important way to deal with quantum uncertainties
\cite{birula1,deutsch,maass}. Recently, noise-disturbance tradeoff
relations have been formulated within the entropic approach
\cite{bhow14}. Entropic uncertainty relations are applied in the
security analysis of quantum key distribution, entanglement
estimation, and quantum communication \cite{ngbw12,cp14}. Entropic
uncertainty relations are reviewed in \cite{ww10,brud11}.

In some respects, however, entropic approach is not considered as
a complete formulation \cite{oppwn10}. Indeed, each entropy is a
function of probability distribution as a whole. Entropic bounds
cannot distinguish the uncertainty inherent in obtaining a
particular combination of the outcomes \cite{oppwn10}. For these
reasons, the so-called fine-grained uncertainty relations have
been introduced \cite{oppwn10}. Although entropic relations are
quite appropriate in many questions, they are only a coarse way of
describing quantum uncertainties. Fine-grained uncertainty
relations for a one-qubit system were examined in \cite{renf13}.
Quantum uncertainties can also be expressed via various majorization
relations, which are currently the subject of active research
\cite{fgg13,prz13,rpz14}.

The aim of the present work is to study fine-grained uncertainty
relations for several measurements in arbitrary finite dimensions.
The paper is organized as follows. In Section \ref{sec2}, the
preliminary material is recalled. In Section \ref{sec3}, we
discuss fine-grained uncertainty relations in general. Explicit
upper bounds are obtained for arbitrary set of mutually unbiased
bases. We extend such bounds for the case of mutually unbiased
measurements recently proposed in \cite{kagour}. We also derive
uncertainty bounds for mutually biased bases. In Section
\ref{sec4}, we develop the proposed method with two measurements
related to state discrimination. The case of three rank-one
projectors is addressed in Section \ref{sec5}. It is exemplified
with mutually unbiased bases in three-dimensional Hilbert space.
In Section \ref{sec6}, we conclude the paper with a summary of
results.

\section{Definitions and notation}\label{sec2}

In this section, preliminary facts are briefly outlined. Let
$\lnp(\hh)$ be the space of linear operators on $d$-dimensional
Hilbert space $\hh$. By $\lsa(\hh)$ and $\lsp(\hh)$, we will
respectively mean the real space of Hermitian operators and the
set of positive ones. For two operators $\ax,\ay\in\lnp(\hh)$,
their Hilbert--Schmidt product is defined as \cite{watrous1}
\begin{equation}
\langle\ax{\,},\ay\rangle_{\mathrm{hs}}:=\tr(\ax^{\dagger}\ay)
\ . \label{axay}
\end{equation}
This inner product induces the norm
$\|\ax\|_{2}=\langle\ax{\,},\ax\rangle_{\mathrm{hs}}^{1/2}$. For
any $\ax\in\lnp(\hh)$, we put $|\ax|\in\lsp(\hh)$ as a unique
positive square root of $\ax^{\dagger}\ax$. The eigenvalues of
$|\ax|$ counted with multiplicities are the singular values of
$\ax$, written $\sigma_{j}(\ax)$. For all real $q\geq1$, the
Schatten $q$-norm is defined as \cite{watrous1}
\begin{equation}
\|\ax\|_{q}:=
\Bigl(\sum\nolimits_{j=1}^{d}\sigma_{j}(\ax)^{q}{\,}\Bigr)^{1/q}
\ . \label{schnd}
\end{equation}
This family gives the trace norm
$\|\ax\|_{1}=\tr|\ax|$ for $q=1$, the Hilbert--Schmidt
norm $\|\ax\|_{2}=\sqrt{\tr(\ax^{\dagger}\ax)}$ for $q=2$, and the
spectral norm
\begin{equation}
\|\ax\|_{\infty}=\max\bigl\{\sigma_{j}(\ax):{\>}1\leq{j}\leq{d}\bigr\}
 \label{spnf}
\end{equation}
for $q=\infty$. Note that the norm (\ref{schnd}) is merely the
vector $q$-norm of the tuple of singular values of $\ax$. For any
operator $\am\in\lsp(\hh)$, the spectral norm coincides with the
maximum of its eigenvalues.

In general, quantum measurements are treated within the POVM
formalism \cite{peresq}. Let $\mc=\{\mm_{j}\}$ be a set of
elements $\mm_{j}\in\lsp(\hh)$, satisfying the completeness
relation
\begin{equation}
\sum\nolimits_{j} \mm_{j}=\pen
\ . \label{cmprl}
\end{equation}
Here, the symbol $\pen$ denotes the identity operator on $\hh$.
The set $\mc=\{\mm_{j}\}$ is a positive operator-valued measure
(POVM). Let the pre-measurement state of a system be described by
density operator $\bro\in\lsp(\hh)$ such that $\tr(\bro)=1$. The
probability of $j$-th outcome is then written as \cite{peresq}
\begin{equation}
p_{j}(\mc|\bro)=\tr(\mm_{j}\bro)
\ . \label{njpr}
\end{equation}
When POVM elements are all orthogonal projectors, we deal with a
projective measurement. Unlike projective measurements, a number
of different outcomes in POVM measurements may exceed
dimensionality of the Hilbert space \cite{peresq}. For all
normalized states $|\psi\rangle\in\hh$, we have
\begin{equation}
\langle\psi|\mm_{j}|\psi\rangle\leq\|\mm_{j}\|_{\infty}
\ . \label{schin1}
\end{equation}
This inequality is saturated with a corresponding eigenvector of
$\mm_{j}$. Combining (\ref{schin1}) with the spectral
decomposition of $\bro$, we further obtain
\begin{equation}
p_{j}(\mc|\bro)\leq\|\mm_{j}\|_{\infty}
\ . \label{nmbn}
\end{equation}
Irrespectively to the pre-measurement state, therefore,
probabilities of the form (\ref{njpr}) can generally be strictly
less than $1$.  Of course, each probability of a projective
measurement can reach the maximum $1$. Discussion of entropic
uncertainty relations for POVMs can be found in \cite{ccyz12}.

Some types of quantum measurements are of special interest
in quantum information processing. Let
$\clb^{(1)}=\bigl\{|b_{j}^{(1)}\rangle\bigr\}$ and
$\clb^{(2)}=\bigl\{|b_{k}^{(2)}\rangle\bigr\}$ be two orthonormal
bases in $d$-dimensional Hilbert space $\hh$. When for all $j$ and
$k$ we have
\begin{equation}
\bigl|\langle{b}_{j}^{(1)}|b_{k}^{(2)}\rangle\bigr|=\frac{1}{\sqrt{d}}
\ , \label{twb}
\end{equation}
these bases are mutually unbiased \cite{bz10}. The set
$\cmb=\left\{\clb^{(1)},\ldots,\clb^{(N)}\right\}$ is a set of
mutually unbiased bases (MUBs), when any two bases from $\cmb$ are
mutually unbiased. For example, three eigenbases of the Pauli
matrices are mutually unbiased. Known constructions for MUBs have
been considered in \cite{wf89,bbrv02,kr04,wb05}. Various
applications of MUBs in quantum theory are reviewed in
\cite{bz10}. It is known that MUBs for higher-dimensional
orbital-angular-momentum states can be used to encode bits of
information in alignment with the BB84 protocol
\cite{bpp00,cbkg02,ylh08}. Experimental studies of such
propositions heve been addressed in \cite{gjvwz06,mdg13}. Entropic
uncertainty relations for $d+1$ mutually unbiased bases in
$d$-dimensional Hilbert space were derived in
\cite{ivan92,sanchez93}. The author of \cite{sanchez93} also gave
the exact bounds for the qubit case $d=2$. These exact relations
have been generalized with use of the Tsallis and R\'{e}nyi
entropies \cite{rastqip,rastctp}. Specific uncertainty relations
for MUBs and some applications were examined in \cite{ballester,MWB10}.

Basic constructions of MUBs are related to the case, when $d$ is
power of a prime number. If the dimensionality is another
composite number, maximal sets of MUBs are an open problem
\cite{bz10}. Hence, we may try to fit ``unbiasedness'' with weaker
conditions. Recently, the authors of \cite{kagour} proposed the
concept of mutually unbiased measurements. Let us consider two
POVM measurements $\mc=\{\mm_{j}\}$ and $\nc=\{\nm_{k}\}$. Each of
them contains $d$ elements such that
\begin{align}
& \tr(\mm_{j})=\tr(\nm_{k})=1
\ , \label{tmn1}\\
& \tr(\mm_{j}\nm_{k})=\frac{1}{d}
\ . \label{dmn1}
\end{align}
The Hilbert--Schmidt product of two elements from the same POVM
depends on the so-called efficiency parameter $\varkappa$
\cite{kagour}. Namely, one has
\begin{equation}
\tr(\mm_{j}\mm_{k})=\delta_{jk}{\,}\varkappa
+(1-\delta_{jk}){\>}\frac{1-\varkappa}{d-1}
\ , \label{mjmk}
\end{equation}
and similarly for the elements of $\nc$. The efficiency parameter
satisfies \cite{kagour}
\begin{equation}
\frac{1}{d}<\varkappa\leq1
\ . \label{vklm}
\end{equation}
For $\varkappa=1/d$ we have the trivial case, in which
$\mm_{j}=\pen/d$ for all $j$. The value $\varkappa=1$, when
possible, leads to the standard case of mutually unbiased bases.
More precise bounds on $\varkappa$ will depend on a construction
of measurement operators. The efficiency parameter shows how close
the measurement operators are to rank-one projectors
\cite{kagour}. For the given $\varkappa$, we take the set
$\mmb=\left\{\mc^{(1)},\ldots,\mc^{(N)}\right\}$ of POVMs
satisfying (\ref{mjmk}). When each two POVMs also obey conditions
of the forms (\ref{tmn1}) and (\ref{dmn1}), the set $\mmb$ is a
set of mutually unbiased measurements (MUMs). If we allow
$\varkappa\neq1$, then one can build $d+1$ MUMs in $d$-dimensional
Hilbert space for arbitrary $d$ \cite{kagour}.

Mutually unbiased bases are used in multi-level schemes of quantum
secret sharing \cite{ylh08}. So-called mutually biased bases form
another important tool for constructing secret-sharing protocols
\cite{ylh08}. By $|x\rangle$, we mean kets of the standard basis
in $d$-dimensional Hilbert space. Let us begin with the unitary
transformation
\begin{equation}
\vm_{F}{\,}|x\rangle=
\frac{1}{\sqrt{d}}{\,}\sum_{y=0}^{d-1}{\exp({\iu}xy\phi){\,}|y\rangle}
\ , \label{ftsb}
\end{equation}
where $\phi=2\pi/d$ and $x\in\{0,\ldots,d-1\}$. In the
computational basis, the Fourier transform acts as (\ref{ftsb}).
To each $t\in\{0,\ldots,d-1\}$, we then assign the set of $d$
vectors \cite{ylh08}
\begin{equation}
|a_{x}^{(t)}\rangle
=\vm_{F}{\,}|x\rangle+\frac{\exp({\iu}t\phi)-1}{\sqrt{d}}{\>}|0\rangle
\ . \label{axtdf}
\end{equation}
The sets $\cla^{(t)}=\bigl\{|a_{x}^{(t)}\rangle\bigr\}$ are all
orthonormal and complete. Following \cite{ylh08}, we call them
mutually biased bases (MBBs). Simple calculations give the
inner product
\begin{equation}
\langle{a}_{x}^{(s)}|a_{y}^{(t)}\rangle
=\delta_{xy}+\frac{{\exp}{\bigl(\iu(t-s)\phi\bigr)}-1}{d}
\ . \label{axy}
\end{equation}
Similarly to MUBs, mutually biased bases can be used in building
effective schemes of quantum secret sharing \cite{ylh08}. In the
next section, we will derive fine-grained uncertainty relations
for MUBs, MUMs, and MBBs.

\section{On fine-grained uncertainty relations in general}\label{sec3}

In this section, we consider the concept of fine-grained
uncertainty relations in principle. A unified approach to their
deriving is then discussed. Further, some general results with an
arbitrary number of unbiased measurements are obtained. The case
of $d$ MBBs is addressed as well. The notion of fine-grained
uncertainty relations has been introduced in \cite{oppwn10}. Some
relations of such a kind for a one-qubit system were examined in
\cite{renf13}.

In some respects, entropies give a coarse way of measuring
uncertainty and incompatibility of a set of quantum measurements
\cite{oppwn10,bm13}. This approach does not distinguish the
uncertainty inherent in obtaining a concrete string of outcomes
for different measurements. Hence, the authors of \cite{oppwn10}
proposed examining inequalities for particular sets of outcomes.
An alternate  approach to measure incompatibility of noncommuting
observables was considered in \cite{bm13}.

Let us consider the corresponding formulation for a set of $N$
MUBs in $d$-dimensional space. In every base, we will number kets
by integers from $0$ up to $d-1$. To each $t=1,\ldots,N$, we
assign an integer $j(t)\in\{0,\ldots,d-1\}$. Assuming equal
weights, we introduce the quantity
\begin{equation}
\underset{\bro}{\max}{\>}\frac{1}{N}{\,}
\sum_{t=1}^{N}{p_{j(t)}(\clb^{(t)}|\bro)}
\ . \label{bnbr}
\end{equation}
Suppose that we have obtained upper bounds on the sum (\ref{bnbr})
for all possible strings $j(1),\ldots,j(N)$. Such upper bounds will
form a fine-grained uncertainty relation for a set of $N$ MUBs. In
general, fine-grained uncertainty relations with unequal weights
may be of interest \cite{oppwn10}. Then upper bounds will also
depend on chosen probabilities of different measurements.

We now consider formulation of fine-grained uncertainty relations
in the case of $N$ generalized  measurements. We should take into
account the fact that particular probabilities may never reach the
value $1$. The unambiguous state discrimination provides one
example of such a kind. Particular probabilities are bounded from
above according to (\ref{nmbn}). For a particular choice of POVM
elements, one per each taken measurement, we herewith take
summarized norms:
\begin{equation}
S:=\sum_{t=1}^{N}{\bigl\|\mm_{j(t)}^{(t)}\bigr\|_{\infty}}
\ . \label{snod}
\end{equation}
To obtain fine-grained uncertainty relations, we aim to estimate
from above all the possible quantities of the form
\begin{equation}
\underset{\bro}{\max}{\>}\frac{1}{S}{\,}
\sum_{t=1}^{N}{p_{j(t)}(\mc^{(t)}|\bro)}
\ . \label{mcbr}
\end{equation}
In this approach, different measurements are assumed to be made
with equal probabilities. Upper bounds on quantities of the form
(\ref{mcbr}) will give a fine-grained uncertainty relation.

Let us consider fine-grained uncertainty relations for MUBs. The
sum under maximization in (\ref{bnbr}) can be represented as
$\tr(\bn\bro)$ in terms of the positive operator
\begin{equation}
\bn=\frac{1}{N}{\,}\sum_{t=1}^{N}{|b_{j}^{(t)}\rangle\langle{b}_{j}^{(t)}|}
\ . \label{posb}
\end{equation}
Here, each index $j$ is prescribed for the given $t$. A simple
bound on the quantity (\ref{bnbr}) follows from the Cauchy--Schwarz
inequality. Applying it to the Hilbert--Schmidt inner product, we
obtain
\begin{equation}
\langle\bn{\,},\bro\rangle_{\mathrm{hs}}\leq
\|\bn\|_{2}{\,}\|\bro\|_{2}
\ . \label{bb22}
\end{equation}
Due to (\ref{twb}), we further calculate the squared
Hilbert--Schmidt norm
\begin{equation}
\|\bn\|_{2}^{2}=\tr(\bn^{\dagger}\bn)=\frac{1}{N^{2}}\left(N+\frac{N^{2}-N}{d}\right)
{\,}. \label{bn22}
\end{equation}
Combining this with (\ref{bb22}) leads to a state-dependent bound
\begin{equation}
\frac{1}{N}{\,}
\sum_{t=1}^{N}{p_{j(t)}(\clb^{(t)}|\bro)}
\leq\sqrt{\tr(\bro^{2})}
{\,}\sqrt{\frac{N+d-1}{Nd}}
{\ }. \label{fgmub00}
\end{equation}
Following \cite{kljr14}, we may regard (\ref{fgmub00}) as a
purity-based uncertainty bound. Purity-based entropic uncertainty
relations were recently considered in \cite{kljr14}. For any pure
state $|\psi\rangle\in\hh$, we further write
\begin{equation}
\frac{1}{N}{\,}
\sum_{t=1}^{N}{p_{j(t)}(\clb^{(t)}|\psi)}
\leq\sqrt{\frac{N+d-1}{Nd}}
{\ }. \label{mub0p}
\end{equation}
This bound also gives a state-independent upper bound, i.e., a
fine-grained uncertainty relation. It turns out that the
uncertainty relation (\ref{mub0p}) can be improved. As
state-independent formulations are of great interest in quantum
information theory, we address this question in more details.
Estimating the spectral norm of (\ref{posb}), the following result
takes place.

\newtheorem{prop1}{Proposition}
\begin{prop1}\label{pan1}
Let $\cmb=\left\{\clb^{(1)},\ldots,\clb^{(N)}\right\}$ be a set of
MUBs in $d$-dimensional Hilbert space $\hh$, and let some integer
$j(t)\in\{0,\ldots,d-1\}$ be assigned to each $t=1,\ldots,N$. For
arbitrary density matrix $\bro$, we have
\begin{equation}
\frac{1}{N}{\,}\sum_{t=1}^{N}{p_{j(t)}(\clb^{(t)}|\bro)}
\leq\frac{1}{d}\left(1+\frac{d-1}{\sqrt{N}}\right)
{\,}. \label{fgmub0}
\end{equation}
\end{prop1}

{\bf Proof.} Similarly to (\ref{nmbn}), we have
$\tr(\bn\bro)\leq\|\bn\|_{\infty}$. In appendix A of the paper
\cite{rast13b}, we have shown the following. For any
$d$-dimensional vector $w$, the corresponding vector norms satisfy
\begin{equation}
\|w\|_{\infty}\leq\frac{1}{d}
\left(\|w\|_{1}+\sqrt{d-1}{\,}\sqrt{d{\,}\|w\|_{2}^{2}-\|w\|_{1}^{2}}\right)
{\,}. \label{wwst}
\end{equation}
By construction, the operator (\ref{posb}) obeys
$\|\bn\|_{1}=\tr(\bn)=1$. Combining this with (\ref{bn22}), we
write the relation
\begin{equation}
d{\,}\|\bn\|_{2}^{2}-\|\bn\|_{1}^{2}=
\frac{d}{N}\left(1+\frac{N-1}{d}\right)-1=\frac{d-1}{N}
\ . \label{bn23}
\end{equation}
Applying (\ref{wwst}) to the singular values of $\bn$, the norm
$\|\bn\|_{\infty}$ is bounded from above by the right-hand side of
(\ref{fgmub0}). $\blacksquare$

The inequality (\ref{fgmub0}) is a fine-grained uncertainty
relation of the state-independent form for an arbitrary set of
MUBs. By construction, this relation is approximate. For a small
number of measurements, exact results may be obtained. Some of
such cases are considered in the next sections. Let us compare the
state-independent bounds (\ref{mub0p}) and (\ref{fgmub0}). Some
calculations show that the right-hand side of (\ref{fgmub0})
cannot exceed (\ref{mub0p}). We now suppose that $d$ is a prime
power and, herewith, there exists a set of $N=d+1$ MUBs. To
compare the bounds (\ref{mub0p}) and (\ref{fgmub0}), we may
calculate a relative difference between them. For instance, this
relative difference is about 6\% for $d=3$, about 13\% for $d=8$,
and it increases with growth of dimensionality. For sufficiently
large $d$, the relative difference tends to $\sqrt{2}-1$ that is
about 40\%. In high dimensions, therefore, the state-independent
bound (\ref{fgmub0}) is significantly better. Thus, our method has
allowed to improve essentially the obvious bound (\ref{mub0p}).
The result (\ref{fgmub0}) can easily be extended to mutually
unbiased measurements. The corresponding relation is formulated as
follows.

\newtheorem{prop2}[prop1]{Proposition}
\begin{prop2}\label{pan2}
Let $\mmb=\left\{\mc^{(1)},\ldots,\mc^{(N)}\right\}$ be a set of
mutually unbiased measurements of the efficiency $\varkappa$ in
$d$-dimensional Hilbert space $\hh$. Let an integer $j(t)$ be
assigned to each $t=1,\ldots,N$. For arbitrary density matrix
$\bro$, we have
\begin{equation}
\frac{1}{N}{\,}
\sum_{t=1}^{N}{p_{j(t)}(\mc^{(t)}|\bro)}
\leq\frac{1}{d}\left(1+\sqrt{\frac{(d-1)(\varkappa{d}-1)}{N}}\right)
{\,}. \label{fgmub1}
\end{equation}
\end{prop2}

{\bf Proof.} We can proceed similarly to the previous proof.
Instead of (\ref{posb}), we consider the positive operator
\begin{equation}
\nq=\frac{1}{N}{\,}\sum_{t=1}^{N}{\mm_{j}^{(t)}}
\ . \label{posb1}
\end{equation}
Due to (\ref{tmn1}), we again have $\|\nq\|_{1}=\tr(\nq)=1$. Using
(\ref{dmn1}) and (\ref{mjmk}), we further calculate the squared
Hilbert--Schmidt norm
\begin{equation}
\|\nq\|_{2}^{2}=\tr(\nq^{\dagger}\nq)=\frac{1}{N^{2}}\left(N\varkappa+\frac{N^{2}-N}{d}\right)
{\,}. \label{bm22}
\end{equation}
Corresponding substitutions immediately gives the upper bound
(\ref{fgmub1}). $\blacksquare$

The inequality (\ref{fgmub1}) gives a fine-grained uncertainty
relation for any set of MUMs. As only conditions of the forms
(\ref{tmn1}) and (\ref{dmn1}) were used in the proof, the bound is
independent of details of construction of measurement elements. In
the case $\varkappa=1$, the result (\ref{fgmub1}) is reduced to
(\ref{fgmub0}). With growth of the efficiency parameter
$\varkappa$, the right-hand side of (\ref{fgmub1}) increases from
$1/d$ up to the right-hand side of (\ref{fgmub0}). Finally, we
note that the Cauchy--Schwarz inequality leads to the
state-dependent relation
\begin{equation}
\frac{1}{N}{\,}
\sum_{t=1}^{N}{p_{j(t)}(\mc^{(t)}|\bro)}
\leq
\sqrt{\tr(\bro^{2})}
{\,}\sqrt{\frac{N+\varkappa{d}-1}{Nd}}
{\ }. \label{mum0}
\end{equation}
Like (\ref{fgmub00}), we merely use here an analog of (\ref{bb22})
and (\ref{bm22}). The inequality (\ref{mum0}) provides a
purity-based form of uncertainty bounds for any set of MUMs. For
pure state $|\psi\rangle\in\hh$, we obtain
\begin{equation}
\frac{1}{N}{\,}
\sum_{t=1}^{N}{p_{j(t)}(\mc^{(t)}|\psi)}
\leq
{\,}\sqrt{\frac{N+\varkappa{d}-1}{Nd}}
{\ }. \label{mum0p}
\end{equation}
It is a state-independent bound, which takes place for all states.
This state-independent bound is sufficiently weaker than
(\ref{fgmub1}). For instance, in the limiting case
$\varkappa\to1/d$ we have the upper bounds $1/d$ from
(\ref{fgmub1}) and only $\sqrt{1/d}$ from (\ref{mum0p}). These two
values considerably differ for large $d$.

In a similar manner, we will obtain a fine-grained uncertainty
relation for $d$ MBBs in $d$-dimensional Hilbert space. Recall
that such bases are used in some schemes of quantum secret sharing
\cite{ylh08}. We have the following statement.

\newtheorem{prop3}[prop1]{Proposition}
\begin{prop3}\label{pan3}
Let $\mma=\left\{\cla^{(0)},\ldots,\cla^{(d-1)}\right\}$ be the
set of $d$ MBBs in $d$-dimensional Hilbert space $\hh$. Let the
given integers $x(0),\ldots,x(d-1)$ form a permutation of the
integers $0,\ldots,d-1$. For arbitrary density matrix $\bro$, we
have
\begin{equation}
\frac{1}{d}{\,}
\sum_{t=0}^{d-1}{p_{x(t)}(\cla^{(t)}|\bro)}
\leq\frac{1}{d}\left(1+\sqrt{2-\frac{2}{d}}{\,}\right)
{\,}. \label{mbb1}
\end{equation}
\end{prop3}

{\bf Proof.} Let us consider the positive operator
\begin{equation}
\am=\frac{1}{d}{\,}\sum_{t=0}^{d-1}{{|a_{x}^{(t)}\rangle\langle{a}_{x}^{(t)}|}}
\ . \label{posa1}
\end{equation}
Here, we have $\|\am\|_{1}=\tr(\am)=1$. The precondition of
the theorem implies $x(s)\neq{x}(t)$ for $s\neq{t}$. Combining
this with (\ref{axy}) then gives
\begin{equation}
d^{2}\|\am\|_{2}^{2}=d+
\frac{1}{d^{2}}
\sum_{{\substack{s,t=0 \\ s\neq{t}}}}^{d-1}
{\left|{\exp}{\bigl(\iu(t-s)\phi\bigr)}-1\right|^{2}}
\ , \label{ahsh1}
\end{equation}
where $\phi=2\pi/d$. Note that the condition $s\neq{t}$ can quite
be ignored in the sum of the right-hand side of (\ref{ahsh1}).
Hence, this sum is calculated as
\begin{equation}
2d^{2}-2\sum_{s,t=0}^{d-1}
{\cos\bigl((t-s)\phi\bigr)}=2d^{2}
\ . \label{ahsh2}
\end{equation}
Indeed, the sum of cosines can be rewritten as the real part of
the sum
\begin{equation}
\sum_{s=0}^{d-1}\exp(-\iu{s}\phi)\sum_{t=0}^{d-1}\exp(\iu{t}\phi)=0
\ . \label{zerosum}
\end{equation}
Therefore, we have
\begin{equation}
d{\,}\|\am\|_{2}^{2}-\|\am\|_{1}^{2}=\frac{2}{d}
\ . \label{ahsh3}
\end{equation}
Using (\ref{wwst}), the spectral norm $\|\am\|_{\infty}$ is
bounded from above by the right-hand side of (\ref{mbb1}).
$\blacksquare$

The inequality (\ref{mbb1}) gives a fine-grained uncertainty
relation of the set of $d$ MBBs. It is instructive to compare this
bound with the upper bound (\ref{fgmub0}) taken with $N=d$ MUBs.
Except for low dimensions, the bound (\ref{fgmub0}) is larger
enough. For large $d$, the bound (\ref{mbb1}) is approximately
${\left(\sqrt{2}+1\right)}/d$, whereas the bound (\ref{fgmub0})
with $N=d$ is approximately $\sqrt{1/d}$. This observation is a
manifestation of the fact that mutually biased bases somehow
deviate from an equipartition provided by mutually unbiased bases.

\section{Relations for two measurements used in state discrimination}\label{sec4}

In this section, we consider fine-grained uncertainty relations
for two measurements related to state discrimination.
Discrimination of quantum states is important in applications
\cite{helstrom,davies} and leading to interesting problems
\cite{ban04}. There exist two basic approaches to discrimination
between non-identical pure states $|\theta_{+}\rangle$ and
$|\theta_{-}\rangle$, which we will parameterize as
\begin{equation}
|\theta_{+}\rangle=
\begin{pmatrix}
\cos\theta \\
\sin\theta
\end{pmatrix}
{\,}, \qquad
|\theta_{-}\rangle=
\begin{pmatrix}
\cos\theta \\
-\sin\theta
\end{pmatrix}
{\,} . \label{etst}
\end{equation}
As the inner product is
$\langle\theta_{+}|\theta_{-}\rangle=\cos2\theta$, we focus our
consideration to the values $2\theta\in(0;\pi/2)$ and, herewith,
to non-identical and non-orthogonal states. Therefore, the value
$\theta=0$ should be left out. In the Helstrom scheme
\cite{helstrom}, the optimal measurement is reached by the set
$\nc=\left\{\nm_{+},\nm_{-}\right\}$ of two projectors
\begin{equation}
\nm_{\pm}=|n_{\pm}\rangle\langle{n}_{\pm}|
\ , \qquad
|n_{\pm}\rangle=\frac{1}{\sqrt{2}}
\begin{pmatrix}
1 \\
\pm1
\end{pmatrix}
{\,}. \label{mpmd}
\end{equation}
In the Helstrom scheme of distinguishing between
$|\theta_{+}\rangle$ and $|\theta_{-}\rangle$, the average
probability of correct answer is optimized \cite{helstrom}. For
equiprobable states, this probability is equal to
\begin{equation}
P_{D}=\frac{1}{2}
\left(
1+\sqrt{1-\bigl|\langle\theta_{+}|\theta_{-}\rangle\bigr|^{2}}
\right)=
\frac{1+\sin2\theta}{2}
\ . \label{coan}
\end{equation}
Of course, the probability (\ref{coan}) becomes maximal for two
orthogonal states.

The second approach known as the unambiguous discrimination is
essential in quantum cryptography \cite{palma}. This approach
sometimes gives an inconclusive answer, but never makes an error
of mis-identification. Such a formulation is purely quantum in
character. The generalized measurement, minimizing the probability
of inconclusive answer, has been built in the papers
\cite{ivan87,dieks,peres1}. In the context of quantum
cryptography, however, another measurement for state
discrimination may be more appropriate \cite{rast0971}. In the B92
scheme \cite{bennett}, the sender encodes bits 0 and 1 into
non-orthogonal states $|\theta_{+}\rangle$ and
$|\theta_{-}\rangle$ respectively. To implement quantum key
distribution, the receiver must distinguish between
$|\theta_{+}\rangle$ and $|\theta_{-}\rangle$ without an error of
mis-identification. Of course, some bits are assumed to be lost.
Let $|m_{j}\rangle$ be vector orthogonal to $|\theta_{j}\rangle$,
i.e., $\langle{m}_{j}|\theta_{j}\rangle=0$ for $j=\pm$. In the
basis $\{|0\rangle,|1\rangle\}$, we have
\begin{equation}
|m_{+}\rangle=
\begin{pmatrix}
\sin\theta \\
-\cos\theta
\end{pmatrix}
{\,}, \qquad
|m_{-}\rangle=
\begin{pmatrix}
\sin\theta \\
\cos\theta
\end{pmatrix}
{\,} . \label{themp}
\end{equation}
In the original version \cite{bennett}, the receiver randomly
measures one of two projectors $|m_{+}\rangle\langle{m}_{+}|$ and
$|m_{-}\rangle\langle{m}_{-}|$. More economical scheme is
described by the three-element POVM
$\mc=\bigl\{\mm_{+},\mm_{-},\mm_{?}\bigr\}$, where
\begin{align}
\mm_{+}&=\frac{1}{1+\cos2\theta}{\>}|m_{-}\rangle\langle{m}_{-}|
{\>}, \quad
\mm_{-}=\frac{1}{1+\cos2\theta}{\>}|m_{+}\rangle\langle{m}_{+}|
{\>}, \nonumber\\
\mm_{?}&=\pen-\mm_{+}-\mm_{-}
{\>}. \label{thpovm}
\end{align}
By construction, we have
$\langle\theta_{-}|\mm_{+}|\theta_{-}\rangle=\langle\theta_{+}|\mm_{-}|\theta_{+}\rangle=0$.
If the POVM measurement $\mc$ has given the outcome ``$+$'', then
the input of discrimination process was certainly
$|\theta_{+}\rangle$; the outcome ``$-$'' implies that the input
was $|\theta_{-}\rangle$. Note that third inconclusive outcome
``$?$'' is allowed, though the problem is effectively
two-dimensional. For the parametrization (\ref{etst}), the
operator $\mm_{?}$ is represented as
\begin{equation}
\mm_{?}=\frac{2\cos2\theta}{1+\cos2\theta}{\>}|0\rangle\langle0|
\ . \label{opin}
\end{equation}
This operator becomes zero with $2\theta=\pi/2$, when the states
(\ref{etst}) are orthogonal. For equiprobable states
$|\theta_{\pm}\rangle$, the average probability of conclusive
answer reads
\begin{equation}
\frac{(\sin2\theta)^{2}}{1+\cos2\theta}=
1-\bigl|\langle\theta_{+}|\theta_{-}\rangle\bigr|
\ . \label{inan}
\end{equation}
This probability vanishes for parallel states and reaches $1$ for
two orthogonal states.

For brevity, we denote the overlap between states
$|\theta_{\pm}\rangle$ by $\eta$, i.e., $\eta=\cos2\theta$. Let us
consider matrices
\begin{equation}
\nm_{\pm}+\mm_{?}=\frac{1}{2}
\begin{pmatrix}
1+2a(\eta) & {\,}\pm1 \\
\pm1 & {\,}1
\end{pmatrix}
{\,}, \qquad
a(\eta):=\frac{2\eta}{1+\eta}
\ . \label{npm1}
\end{equation}
By usual calculations, we then obtain
\begin{equation}
\left\|\nm_{\pm}+\mm_{?}\right\|_{\infty}=
\frac{1}{2}\left(1+a(\eta)+\sqrt{1+a(\eta)^{2}}\right)
{\,}. \label{npm2}
\end{equation}
This value can be reached with the corresponding eigenvectors of
the matrices $\nm_{\pm}+\mm_{?}$. Using
$\|\mm_{?}\|_{\infty}=a(\eta)$, the exact uncertainty bound is
written as
\begin{equation}
\underset{\bro}{\max}{\>}\frac{p_{\pm}(\nc|\bro)+p_{{\,}?}(\mc|\bro)}{1+\|\mm_{?}\|_{\infty}}
=\frac{1}{2}\left(1+\frac{\sqrt{1+2\eta+5\eta^{2}}}{1+3\eta}\right)
{\,}. \label{npm3}
\end{equation}
The right-hand side of (\ref{npm3}) is the upper bound on rescaled
sum of two particular probabilities. This bound decreases from the
value $1$ for $\eta=0$ (orthogonal states) up to the infimum
$(2+\sqrt{2})/4\approx0.854$ for $\eta\to1^{-}$ (parallel states).

We will now consider the matrices $\nm_{+}+\mm_{+}$ and
$\nm_{-}+\mm_{-}$. In the chosen basis, they are respectively
written as
\begin{equation}
\frac{1}{2(1+\eta)}
\begin{pmatrix}
2 & \pm(1+\eta+\sin2\theta) \\
\pm(1+\eta+\sin2\theta) & 2(1+\eta)
\end{pmatrix}
{\,}. \label{nmmn}
\end{equation}
Calculating eigenvalues, we then obtain
\begin{equation}
\left\|\nm_{+}+\mm_{+}\right\|_{\infty}=\left\|\nm_{-}+\mm_{-}\right\|_{\infty}=
\frac{2+\eta+\sqrt{\eta^{2}+4(1+\eta)P_{D}}}{2(1+\eta)}
{\ }. \label{nmmn2}
\end{equation}
Recall that $P_{D}$ denotes the right-hand side of (\ref{coan}).
Combining (\ref{nmmn2}) with
$\|\mm_{\pm}\|_{\infty}=(1+\eta)^{-1}$ gives the exact uncertainty
bound
\begin{equation}
\underset{\bro}{\max}{\>}\frac{p_{j}(\nc|\bro)+p_{j}(\mc|\bro)}{1+\|\mm_{j}\|_{\infty}}=
\frac{1}{2}\left(1+\frac{\sqrt{\eta^{2}+4(1+\eta)P_{D}}}{2+\eta}\right)
{\,}, \label{nmmn3}
\end{equation}
where $j=\pm$. The bound (\ref{nmmn3}) decreases from the value
$1$ for $\eta=0$ (orthogonal states) up to the infimum
$1/2+\sqrt{5}/6\approx0.873$ for $\eta\to1^{-}$ (parallel states).

Finally, we consider the matrices $\nm_{+}+\mm_{-}$ and
$\nm_{-}+\mm_{+}$. These matrices are respectively expressed as
\begin{equation}
\frac{1}{2(1+\eta)}
\begin{pmatrix}
2 & \pm(1+\eta-\sin2\theta) \\
\pm(1+\eta-\sin2\theta) & 2(1+\eta)
\end{pmatrix}
{\,}. \label{mnmn}
\end{equation}
By calculations, we  further obtain
\begin{equation}
\left\|\nm_{+}+\mm_{-}\right\|_{\infty}=\left\|\nm_{-}+\mm_{+}\right\|_{\infty}=
\frac{2+\eta+\sqrt{\eta^{2}+4(1+\eta)P_{E}}}{2(1+\eta)}
{\ }. \label{mnmn2}
\end{equation}
Here, the term $P_{E}=1-P_{D}$ is the error probability in the
Helstrom scheme. The exact uncertainty bound reads
\begin{equation}
\underset{\bro}{\max}{\>}\frac{p_{j}(\nc|\bro)+p_{k}(\mc|\bro)}{1+\|\mm_{k}\|_{\infty}}=
\frac{1}{2}\left(1+\frac{\sqrt{\eta^{2}+4(1+\eta)P_{E}}}{2+\eta}\right)
{\,}, \label{mnmn3}
\end{equation}
where the pair $(j,k)$ is either $(+,-)$ or $(-,+)$. Unlike
(\ref{nmmn3}), the bound (\ref{mnmn3}) increases with growing
$\eta$. The bound (\ref{mnmn3}) is equal to $1/2$ for $\eta=0$
(orthogonal states) and tends to $1/2+\sqrt{5}/6\approx0.873$ in
the limit $\eta\to1^{-}$ (parallel states). Note also that the
quantities (\ref{nmmn3}) and (\ref{mnmn3}) coincide in this limit.
It is a manifestation of the fact that states become
indistinguishable. On the other hand, in the case of orthogonal
states the quantity (\ref{nmmn3}) twice exceeds (\ref{mnmn3}). In
combination, the formulas (\ref{npm3}), (\ref{nmmn3}), and
(\ref{mnmn3}) give an exact fine-grained uncertainty relation for
two measurements used in quantum state discrimination.

\section{Relations for three rank-one projectors in $d$-dimensional space}\label{sec5}

In this section, we will derive fine-grained uncertainty relations
for three MUBs in a finite-dimensional Hilbert space. The case of
two rank-one projectors is effectively two-dimensional. Three
projectors generally lead to a situation that cannot be reduced in
such a way. We will also study a concrete example of certain
physical interest. Our approach is based on calculation of the
maximal eigenvalue of a certain positive matrix. This matrix is
expressed as the sum of three rank-one projectors. Let
$|b_{1}\rangle$, $|b_{2}\rangle$, and $|b_{3}\rangle$ be three
unit vectors such that pairwise overlaps are all equal to
$\eta<1$, i.e., $\bigl|\langle{b}_{j}|b_{k}\rangle\bigr|=\eta$ for
$j\neq{k}$. We aim to find the spectral norm of the positive
operator
\begin{equation}
|b_{1}\rangle\langle{b}_{1}|+|b_{2}\rangle\langle{b}_{2}|+|b_{3}\rangle\langle{b}_{3}|
\ . \label{amb123}
\end{equation}
When the vectors are taken from the MUBs $\clb^{(1)}$,
$\clb^{(2)}$, $\clb^{(3)}$, i.e., $|b_{j}\rangle\in\clb^{(j)}$, we
have $\eta=d^{-1/2}$. We will assume that the subspace
${\mathrm{span}}\bigl\{|b_{1}\rangle,|b_{2}\rangle,|b_{3}\rangle\bigr\}$
is three-dimensional, though our approach is quite applicable in
two actual dimensions. Of course, the two-dimensional case is
easier to analyze.

We begin with three vectors, for which inner products are all
real. First, we introduce a useful parametrization of involved
vectors. Let us consider the sub-normalized vectors
\begin{align}
|b_{1}^{\perp}\rangle&:=|b_{1}\rangle-|b_{3}\rangle\langle{b}_{3}|b_{1}\rangle
\ , \label{prp1}\\
|b_{2}^{\perp}\rangle&:=|b_{2}\rangle-|b_{3}\rangle\langle{b}_{3}|b_{2}\rangle
\ . \label{prp2}
\end{align}
In the case considered, the inner product
$\langle{b}_{1}^{\perp}|b_{2}^{\perp}\rangle$ can be assumed to be
positive real. In the subspace
${\mathrm{span}}\bigl\{|b_{1}^{\perp}\rangle,|b_{2}^{\perp}\rangle\bigr\}$,
we take orthonormal basis such that the first ket is proportional
to $|b_{1}^{\perp}\rangle+|b_{2}^{\perp}\rangle$ and the second
ket is proportional to
$|b_{2}^{\perp}\rangle-|b_{1}^{\perp}\rangle$. The third auxiliary
ket is merely equal to $|b_{3}\rangle$. Due to the defined
auxiliary kets, we will deal with the following vectors:
\begin{equation}
|c_{1}\rangle=
\begin{pmatrix}
\alpha \\
-\beta \\
\eta
\end{pmatrix}
{\,} , \qquad
|c_{2}\rangle=
\begin{pmatrix}
\alpha \\
\beta \\
\eta
\end{pmatrix}
{\,} , \qquad
|c_{3}\rangle=
\begin{pmatrix}
0 \\
0 \\
1
\end{pmatrix}
{\,} . \label{b123}
\end{equation}
For the given $\eta$, the positive real numbers $\alpha$ and
$\beta$ satisfy
\begin{align}
\alpha^{2}+\beta^{2}+\eta^{2}&=1
\ , \label{nomz}\\
\alpha^{2}-\beta^{2}+\eta^{2}&=\eta
\ . \label{noin}
\end{align}
After calculations, we obtain
\begin{align}
\alpha=\sqrt{\frac{1+\eta-2\eta^{2}}{2}}
\ , \qquad
\beta=\sqrt{\frac{1-\eta}{2}}
\ . \label{y2df}
\end{align}
The sum of three projectors then reads
\begin{equation}
|c_{1}\rangle\langle{c}_{1}|+|c_{2}\rangle\langle{c}_{2}|+|c_{3}\rangle\langle{c}_{3}|=
\begin{pmatrix}
2\alpha^{2} & 0 & 2\alpha\eta \\
0 & 2\beta^{2} & 0 \\
2\alpha\eta & 0 & 2\eta^{2}+1
\end{pmatrix}
{\,} . \label{amuph0}
\end{equation}
The characteristic equation of the matrix (\ref{amuph0}) is
written as
\begin{equation}
(\lambda-2\beta^{2})\bigl(\lambda^{2}-(2+\eta)\lambda+2\alpha^{2}\bigr)=0
\ . \label{chra}
\end{equation}
By calculations, we finally obtain the eigenvalue $1-\eta$ with
multiplicity $2$ and the eigenvalue $1+2\eta$. Thus, the spectral
norm of (\ref{amuph0}) is equal to $1+2\eta$. Hence, we write an
exact uncertainty bound
\begin{equation}
\underset{\bro}{\max}{\>}\frac{1}{3}{\,}
\sum_{j=1}^{3}{\langle{c}_{j}|\bro|c_{j}\rangle}
=\frac{1+2\eta}{3}
\ . \label{12et}
\end{equation}
This relation is tight in the following sense. It is always
saturated with the corresponding eigenvector of (\ref{amuph0}),
when $\bro=|\psi\rangle\langle\psi|$ and
\begin{equation}
|\psi\rangle=
\frac{1}{\sqrt{3(1+2\eta)}}
\begin{pmatrix}
\sqrt{2(1+\eta-2\eta^{2})} \\
0 \\
1+2\eta
\end{pmatrix}
{\,}. \label{crps}
\end{equation}
This form is related to those basis that is used in the
representation (\ref{b123}). The state
(\ref{crps}) has the same overlap with each $|c_{j}\rangle$. This
overlap is equal to the squared root of the right-hand side of
(\ref{12et}). The right-hand side of (\ref{12et}) increases from
the value $1/3$ for $\eta=0$ up to the value $1$ for $\eta=1$. The
former corresponds to three mutually orthogonal projectors,
whereas the latter actually deals with one and the same projector.

We now consider a more complicated example. It is related to the
four MUBs in three-dimensional Hilbert space. In principle, we
may build some parametrization similarly to (\ref{b123}). It will
include an additional phase factor. With explicitly given vectors,
however, direct calculations are more appropriate. By
$\omega=\exp(\iu2\pi/3)$, we mean a primitive root of the
unit. The normalized vectors of these bases are written as
\begin{align}
&\left\{
\begin{pmatrix}
1 \\
0 \\
0
\end{pmatrix},
{\>}
\begin{pmatrix}
0 \\
1 \\
0
\end{pmatrix},
{\>}
\begin{pmatrix}
0 \\
0 \\
1
\end{pmatrix}
\right\}
{\,},
&\left\{
\frac{1}{\sqrt{3}}
\begin{pmatrix}
1 \\
1 \\
1
\end{pmatrix},
{\>}
\frac{1}{\sqrt{3}}
\begin{pmatrix}
1 \\
\omega^{*} \\
\omega
\end{pmatrix},
{\>}
\frac{1}{\sqrt{3}}
\begin{pmatrix}
1 \\
\omega \\
\omega^{*}
\end{pmatrix}
\right\}
{\,}, \label{bas12}\\
&\left\{
\frac{1}{\sqrt{3}}
\begin{pmatrix}
1 \\
\omega \\
1
\end{pmatrix},
{\>}
\frac{1}{\sqrt{3}}
\begin{pmatrix}
1 \\
1 \\
\omega
\end{pmatrix},
{\>}
\frac{1}{\sqrt{3}}
\begin{pmatrix}
\omega \\
1 \\
1
\end{pmatrix}
\right\}
{\,},
&\left\{
\frac{1}{\sqrt{3}}
\begin{pmatrix}
1 \\
1 \\
\omega^{*}
\end{pmatrix},
{\>}
\frac{1}{\sqrt{3}}
\begin{pmatrix}
\omega^{*} \\
1 \\
1
\end{pmatrix},
{\>}
\frac{1}{\sqrt{3}}
\begin{pmatrix}
1 \\
\omega^{*} \\
1
\end{pmatrix}
\right\}
{\,}. \label{bas34}
\end{align}
Here, the symbol $*$ denotes complex conjugation. When one of MUBs
is taken as the standard base, other MUBs can be described in
terms of complex Hadamard matrices. This fact has been used for
classification of MUBs in low dimensions \cite{bwb10}.

The four MUBs (\ref{bas12})--(\ref{bas34}) give eigenbases of the
four operators from the Weyl--Heisenberg group. They were applied
in studies of complementarity of spin-$1$ observables
\cite{kkc09}. These bases were also used in experimental study of
higher-dimensional quantum key distribution protocols based on
mutually unbiased bases \cite{mdg13}. The two bases (\ref{bas12})
are respectively the eigenbases of the generalized Pauli operators
\begin{equation}
\az=\begin{pmatrix}
1 & 0 & 0 \\
0 & \omega & 0 \\
0 & 0 & \omega^{*}
\end{pmatrix}
{\,},
\qquad
\ax=\begin{pmatrix}
0 & 0 & 1 \\
1 & 0 & 0 \\
0 & 1 & 0
\end{pmatrix}
{\,}. \label{azmatr}
\end{equation}
Further, the two bases (\ref{bas34}) are the eigenbases of the
operators $\az\ax$ and $\az\ax^{2}$. The vectors in each base are
arranged according to the order of eigenvalues $1$, $\omega$,
$\omega^{*}$. Except for own eigenbasis, the action of each of the
four operators $\az$, $\ax$, $\az\ax$, $\az\ax^{2}$ leads to
cyclic permutations in other three bases, sometimes with
additional phase factors.

We consider fine-grained uncertainty relations for the projectors
$|b_{1}\rangle\langle{b}_{1}|$, $|b_{2}\rangle\langle{b}_{2}|$,
and $|b_{3}\rangle\langle{b}_{3}|$, where the used vectors are
separately taken from three of the above bases
(\ref{bas12})--(\ref{bas34}). Of course, no two vectors are taken
from one and the same base. First, we consider the vectors
\begin{equation}
|b_{1}\rangle=\frac{1}{\sqrt{3}}
\begin{pmatrix}
 1 \\
 1 \\
 1
\end{pmatrix}
{\,}, \qquad
|b_{2}\rangle=\frac{1}{\sqrt{3}}
\begin{pmatrix}
 1 \\
 \omega \\
 1
\end{pmatrix}
{\,}, \qquad
|b_{3}\rangle=\frac{1}{\sqrt{3}}
\begin{pmatrix}
 1 \\
 \omega^{*} \\
 1
\end{pmatrix}
{\,}. \label{fb1b2b3}
\end{equation}
We easily calculate the sum of projectors
\begin{equation}
|b_{1}\rangle\langle{b}_{1}|+|b_{2}\rangle\langle{b}_{2}|+|b_{3}\rangle\langle{b}_{3}|=
\begin{pmatrix}
1 & 0 & 1 \\
0 & 1 & 0 \\
1 & 0 & 1
\end{pmatrix}
{\,} . \label{amuph01}
\end{equation}
This matrix has the trace $3$ and the determinant $0$. The sum of
second-order principal minors is equal to $2$. So, the
characteristic equation is written as
\begin{equation}
\lambda^{3}-3\lambda^{2}+2\lambda=0
{\ } , \label{cheq01}
\end{equation}
whence the eigenvalues $0$, $1$, $2$ are found. Thus, we obtain
the upper bound
\begin{equation}
\underset{\bro}{\max}{\>}\frac{1}{3}{\,}
\sum_{j=1}^{3}{\langle{b}_{j}|\bro|b_{j}\rangle}
=\frac{2}{3}
\ . \label{013et}
\end{equation}
It is always saturated with the corresponding eigenvector of the
matrix (\ref{amuph01}), when $\bro=|\psi\rangle\langle\psi|$ and
\begin{equation}
|\psi\rangle=
\frac{1}{\sqrt{2}}
\begin{pmatrix}
1 \\
0 \\
1
\end{pmatrix}
{\,}. \label{crps01}
\end{equation}
The relation (\ref{013et}) also holds for some other three-vector
combinations such that no vectors are taken from one and the same
base. For other combinations, the maximizing vector will differ
from (\ref{crps01}). Of course, it can easily be calculated with
each concrete choice of three vectors.

Second, we consider the following unit vectors:
\begin{equation}
|a_{1}\rangle=
\begin{pmatrix}
 0 \\
 0 \\
 1
\end{pmatrix}
{\,}, \qquad
|a_{2}\rangle=
\frac{1}{\sqrt{3}}
\begin{pmatrix}
 1 \\
 \omega \\
 1
\end{pmatrix}
{\,}, \qquad
|a_{3}\rangle=
\frac{1}{\sqrt{3}}
\begin{pmatrix}
 1 \\
 \omega^{*} \\
 1
\end{pmatrix}
{\,}. \label{b1b2b3}
\end{equation}
We should calculate the spectral norm of the matrix
\begin{equation}
|a_{1}\rangle\langle{a}_{1}|+
|a_{2}\rangle\langle{a}_{2}|+|a_{3}\rangle\langle{a}_{3}|=
\frac{1}{3}
\begin{pmatrix}
2 & -1 & 2 \\
-1 & 2 & -1 \\
2 & -1 & 5
\end{pmatrix}
{\,}. \label{amuph1}
\end{equation}
This matrix has the trace $3$ and the determinant $1/3$. The sum
of second-order principal minors is equal to $2$. Then the
characteristic equation reads
\begin{equation}
\lambda^{3}-3\lambda^{2}+2\lambda-\frac{1}{3}=0
\ . \label{cheq3}
\end{equation}
Solving this equation, we find the maximal eigenvalue
\begin{equation}
\max\lambda=1+\frac{2}{\sqrt{3}}{\,}\cos\frac{\pi}{18}\approx2.137
\ . \label{malm}
\end{equation}
Thus, we obtain an upper bound on the sum of three probabilities,
namely
\begin{equation}
\underset{\bro}{\max}{\>}\frac{1}{3}{\,}
\sum_{j=1}^{3}{\langle{a}_{j}|\bro|a_{j}\rangle}
=\frac{1}{3}
\left(
1+\frac{2}{\sqrt{3}}{\,}\cos\frac{\pi}{18}
\right)\approx\frac{2.137}{3}
{\,}. \label{13et}
\end{equation}
Some inspection shows that each combination of three unit vectors
from a triple of MUBs of the set (\ref{bas12})--(\ref{bas34})
gives either (\ref{cheq01}) or (\ref{cheq3}). The former
corresponds to the determinant $0$, the latter corresponds to the
determinant $1/3$. Thus, the relations (\ref{013et}) and
(\ref{13et}) together give a fine-grained uncertainty relation for
the above MUBs in $3$-dimensional Hilbert space. Both the upper
bounds (\ref{013et}) or (\ref{13et}) are strictly less than $1$.
They reflect a complementarity of spin-$1$ observables. Using the
quantities (\ref{013et}) and (\ref{13et}), we may estimate a
quality of the bound of Proposition \ref{pan1}. Substituting
$d=N=3$ into the right-hand side of (\ref{fgmub0}), we obtain the
number ${\left(1+2/\sqrt{3}\right)}/3\approx2.155/3$. It is close
to (\ref{013et}) and very close to (\ref{13et}) from above. This
gives an evidence for the fact that the approximate bound
(\ref{fgmub0}) is good enough, at least for low dimensions.

As was shown in \cite{bwb10}, triples of MUBs in $d=3$ are all
equivalent in the sense of several kinds of transformations
applied to a string of Hadamard matrices. The set of used
transformations includes an overall unitary matrix applied from
the left, diagonal unitary matrices applied from the right,
permutations of vectors within each basis, pairwise exchanges of
two bases, and an overall complex conjugation. It can be checked that
such operations will again lead to either (\ref{013et}) or
(\ref{13et}). For example, multiplication by a diagonal unitary
matrix from the right merely attaches phase factors to each column
of the transformed matrix \cite{bwb10}. Thus, these bounds cover
all fine-grained uncertainty relations for triplets of MUBs in the
case $d=3$. The above method could be applied to triples of MUBs
in other dimensions, but a situation becomes more complicated. For
instance, the authors of \cite{bwb10} found a three-parameter
family of triples in dimension four and two inequivalent triples
in dimension five.

\section{Conclusion}\label{sec6}

We have examined fine-grained uncertainty relations in the case of
several quantum measurements. Formulation of fine-grained
uncertainty relations for generalized measurements is first
considered. Our approach is based on evaluation of spectral norms
of the corresponding positive matrices. In discussed examples, we
assume that different measurements have equal weights. However,
the considered approach is also appropriate for cases, when
measurements of interest have different probabilities. Explicit
upper bounds are obtained for arbitrary set of mutually unbiased
bases and then extended for the case of mutually unbiased
measurements. In a similar manner, we considered uncertainty
relations for so-called mutually biased bases essential in some
schemes of quantum secret sharing. To illustrate an approach to
generalized measurements, we analyzed a fine-grained  uncertainty
relation for two measurements used in state discrimination. One of
them known as the unambiguous state discrimination is closely
related to the B92 protocol of quantum cryptography. We also
examined fine-grained uncertainty relations for three rank-one
projective measurements. As an example, we further addressed the
case of mutually unbiased bases in three-dimensional Hilbert
space. Considered fine-grained uncertainty relations may be useful
in studying secret-sharing and quantum-key-distribution protocols
based on mutually biased or unbiased bases.

\end{document}